\documentclass{article}
\usepackage{fullpage}
\usepackage[american]{babel}
\usepackage{amssymb}
\usepackage{algorithmic}
\usepackage{algorithm}
\usepackage{multirow}
\usepackage{graphicx}

\newcommand{\eqn}[1]{(\ref{#1})}
\newcommand{\beql}[1]{\begin{equation}\label{#1}}
\newcommand{\eeq}{\end{equation}}
\newtheorem{theo}{Theorem}
\newtheorem{defi}{Definiton}

\title{An Efficient Dynamic Programming Algorithm for STR-IC-SEQ-EC-LCS Problem}
\author{Daxin Zhu, Lei Wang, Yingjie Wu, and Xiaodong Wang}

\begin{document}
\maketitle

\begin{abstract}
In this paper, we consider a generalized longest common subsequence problem, in which a constraining sequence of length $s$ must be included as a substring and the other constraining sequence of length $t$ must be excluded as a subsequence of two main sequences and the length of the result must be maximal. For the two input sequences $X$ and $Y$ of lengths $n$ and $m$, and the given two constraining sequences of length $s$ and $t$, we present an $O(nmst)$ time dynamic programming algorithm for solving the new generalized longest common subsequence problem.
The time complexity can be reduced further to cubic time in a more detailed analysis.
The correctness of the new algorithm is proved.
\end{abstract}

\section{Introduction}
The longest common subsequence (LCS) problem is a well-known measurement for computing the similarity of two strings. It can be widely applied in diverse areas, such as file comparison, pattern matching and computational biology\cite{3,4,8,9}.

Given two sequences $X$ and $Y$, the longest common subsequence (LCS) problem is to find a subsequence of $X$ and $Y$ whose length is the longest among all common subsequences of the two given sequences.

For some biological applications some constraints must be applied to the LCS problem. These kinds of variant of the LCS problem are called the constrained LCS (CLCS) problem. Recently, Chen and Chao\cite{1} proposed the more generalized forms of the CLCS problem, the generalized constrained longest common subsequence (GC-LCS) problem.
For the two input sequences $X$ and $Y$ of lengths $n$ and $m$,respectively, and a constraint string $P$ of length $r$, the GC-LCS problem is a set of four problems which are to find the LCS of $X$ and $Y$ including/excluding $P$ as a subsequence/substring, respectively.

In this paper, we consider a more general constrained longest common subsequence problem called STR-IC-SEQ-EC-LCS, in which a constraining sequence of length $s$ must be included as a substring and the other constraining sequence of length $t$ must be excluded as a subsequence of two main sequences and the length of the result must be maximal. We will present the first efficient dynamic programming algorithm for solving this problem.

The organization of the paper is as follows.

In the following 4 sections we describe our presented dynamic programming algorithm for the STR-IC-SEQ-EC-LCS problem.

In section 2 the preliminary knowledge for presenting our algorithm for the STR-IC-SEQ-EC-LCS problem is discussed.
In section 3 we give a new dynamic programming solution for the STR-IC-SEQ-EC-LCS problem with time complexity $O(nmst)$, where $n$ and $m$ are the lengths of the two given input strings, and $s$ and $t$ the lengths of the two constraining sequences.
In section 4 the time complexity is further improved to $O(nmt)$.
Some concluding remarks are in section 5.

\section{Characterization of the STR-IC-SEQ-EC-LCS problem}
A sequence is a string of characters over an alphabet $\sum$. A subsequence of a sequence $X$ is obtained by deleting zero or more characters from $X$ (not necessarily contiguous). A substring of a sequence $X$ is a subsequence of successive characters within $X$.

For a given sequence $X=x_1x_2\cdots x_n$ of length $n$, the $i$th character of $X$ is denoted as $x_i \in \sum$ for any $i=1,\cdots,n$. A substring of $X$ from position $i$ to $j$ can be denoted as $X[i:j]=x_ix_{i+1}\cdots x_j$. If $i\neq 1$ or $j\neq n$, then the substring $X[i:j]=x_ix_{i+1}\cdots x_j$ is called a proper substring of $X$. A substring $X[i:j]=x_ix_{i+1}\cdots x_j$ is called a prefix or a suffix of $X$ if $i=1$ or $j=n$, respectively.

An appearance of sequence $X=x_1x_2\cdots x_n$ in sequence $Y=y_1y_2\cdots y_m$, for any $X$ and $Y$, starting at position $j$ is a sequence of strictly increasing indexes $i_1,i_2,\cdots,i_n$ such that $i_1=j$, and $X=y_{i_1},y_{i_2},\cdots,y_{i_n}$. A compact appearance of $X$ in $Y$ starting at position $j$ is the appearance of the smallest last index $i_n$.
A match for sequences $X$ and $Y$ is a pair $(i,j)$ such that $x_i=y_j$. The total number of matches for $X$ and $Y$ is denoted by $\delta$. It is obvious that $\delta\leq nm$.

For the two input sequences $X=x_1x_2\cdots x_n$ and $Y=y_1y_2\cdots y_m$ of lengths $n$ and $m$, respectively, and two constrained sequences $P=p_1p_2\cdots p_s$ and $Q=q_1q_2\cdots q_t$ of lengths $s$ and $t$, the SEQ-IC-STR-IC-LCS problem is to find a constrained LCS of $X$ and $Y$ including $P$ as a substring and excluding $Q$ as a subsequence.

\begin{defi}\label{df1}
Let $Z(i,j,k,r)$ denote the set of all LCSs of $X[1:i]$ and $Y[1:j]$ such that for each $z\in Z(i,j,k,r)$, $z$ includes $P[1:k]$ as a substring, and excludes $Q[1:r]$ as a subsequence, where $1\leq i\leq n, 1\leq j\leq m, 0\leq k\leq s$, and $0\leq r\leq t$.
The length of an LCS in $Z(i,j,k,r)$ is denoted as $g(i,j,k,r)$.
\end{defi}

\begin{defi}\label{df2}
Let $W(i,j,k,r)$ denote the set of all LCSs of $X[1:i]$ and $Y[1:j]$ such that for each $w\in W(i,j,k,r)$, $w$ excludes $Q[1:r]$ as a subsequence, and includes $P[1:k]$ as a suffix, where $1\leq i\leq n, 1\leq j\leq m, 0\leq k\leq s$, and $0\leq r\leq t$.
The length of an LCS in $W(i,j,k,r)$ is denoted as $f(i,j,k,r)$.
\end{defi}

\begin{defi}\label{df3}
Let $U(i,j,k)$ denote the set of all LCSs of $X[i:n]$ and $Y[j:m]$ such that for each $u\in U(i,j,k)$, $u$ excludes $Q[k:t]$ as a subsequence, where $1\leq i\leq n, 1\leq j\leq m, 0\leq k\leq t$.
The length of an LCS in $U(i,j,k)$ is denoted as $h(i,j,k)$.
\end{defi}

\begin{defi}\label{df4}
Let $V(i,j,k)$ denote the set of all LCSs of $X[1:i]$ and $Y[1:j]$ such that for each $v\in V(i,j,k)$, $v$ excludes $Q[1:k]$ as a subsequence, where $1\leq i\leq n, 1\leq j\leq m, 0\leq k\leq t$.
The length of an LCS in $V(i,j,k)$ is denoted as $v(i,j,k)$.
\end{defi}

The following theorem characterizes the structure of an optimal solution based on optimal solutions to subproblems, for computing the LCSs in $W(i,j,k,r)$, for any $1\leq i\leq n, 1\leq j\leq m, 0\leq k\leq s$, and $0\leq r\leq t$.

\begin{theo}\label{th1}
If $Z[1:l]=z_1,z_2,\cdots,z_l\in W(i,j,k,r)$, then the following conditions hold:
\begin{enumerate}
\item If $i,j,k>0,r=1$, $x_i=y_j=p_k=q_r$, then $z_l\neq x_i$ and $Z[1:l]\in W(i-1,j-1,k,r)$.
\item If $i,j,k>0,r>1$, $x_i=y_j=p_k=q_r$, then $z_l\neq x_i$ implies $Z[1:l]\in W(i-1,j-1,k,r)$; $z_l=x_i$ implies $Z[1:l-1]\in W(i-1,j-1,k-1,r-1)$.
\item If $i,j,k>0$, $x_i=y_j=p_k$ and $r>0, x_i\neq q_r$ or $r=0$, then $z_l=x_i=y_j=p_k$ and $Z[1:l-1]\in W(i-1,j-1,k-1,r)$.
\item If $i,j,k>0$, $x_i=y_j$ and $x_i\neq p_k$, then $z_l\neq x_i$ and $Z[1:l]\in W(i-1,j-1,k,r)$.

\item If $i,j>0,k=0,r=1$, $x_i=y_j=q_r$, then $z_l\neq x_i$ and $Z[1:l]\in W(i-1,j-1,k,r)$.
\item If $i,j>0,k=0,r>1$, $x_i=y_j=q_r$, then $z_l\neq x_i$ implies $Z[1:l]\in W(i-1,j-1,k,r)$; $z_l=x_i$ implies $Z[1:l-1]\in W(i-1,j-1,k,r-1)$.
\item If $i,j>0,k=0$, $x_i=y_j$ and $r>0, x_i\neq q_r$ or $r=0$, then $z_l=x_i$ and $Z[1:l-1]\in W(i-1,j-1,k,r)$.

\item If $i,j>0, x_i\neq y_j$, then $z_l\neq x_i$ implies $Z[1:l]\in W(i-1,j,k,r)$.
\item If $i,j>0, x_i\neq y_j$, then $z_l\neq y_j$ implies $Z[1:l]\in W(i,j-1,k,r)$.
\end{enumerate}
\end{theo}

\noindent{\bf Proof.}

1. In this case, if $x_i=z_l$, then $Z[1:l]$ includes $Q[1:r]$, a contradiction. Therefore, we have $x_i\neq z_l$, and $Z[1:l]$ must be an LCS of $X[1:i-1]$ and $Y[1:j-1]$ including $P[1:k]$ as a suffix and excluding $Q[1:r]$ as a subsequence, i.e. $Z[1:l]\in W(i-1,j-1,k,r)$.

2. There are two subcases to be distinguished in this case.

2.1. If $z_l=x_i$, then $Z[1:l-1]$ is a common subsequence of $X[1:i-1]$ and $Y[1:j-1]$ including $P[1:k-1]$ as a suffix and excluding $Q[1:r-1]$ as a subsequence. We can show that $Z[1:l-1]$ is an LCS of $X[1:i-1]$ and $Y[1:j-1]$ including $P[1:k-1]$ as a suffix and excluding $Q[1:r-1]$ as a subsequence.
Assume by contradiction that there exists a common subsequence $a$ of $X[1:i-1]$ and $Y[1:j-1]$ including $P[1:k-1]$ as a suffix and excluding $Q[1:r-1]$ as a subsequence, whose length is greater than $l-1$. Then the concatenation of $a$ and $z_l$ will result in a common subsequence of $X[1:i]$ and $Y[1:j]$ including $P[1:k]$ as a suffix and excluding $Q[1:r]$ as a subsequence, whose length is greater than $l$. This is a contradiction.
Therefore, in this case we have $Z[1:l-1]\in W(i-1,j-1,k-1,r-1)$.

2.2. If $z_l\neq x_i$, then $Z[1:l]$ must be an LCS of $X[1:i-1]$ and $Y[1:j-1]$ including $P[1:k]$ as a suffix and excluding $Q[1:r]$ as a subsequence, i.e. $Z[1:l]\in W(i-1,j-1,k,r)$.

3. In this case, we have no constraints on $Q$, provided $r>0, x_i\neq q_r$ or $r=0$. Therefore we have $x_i=y_j=p_k=z_l$. It is obvious that $Z[1:l-1]$ is a common subsequence of $X[1:i-1]$ and $Y[1:j-1]$
including $P[1:k-1]$ as a suffix and excluding $Q[1:r]$ as a subsequence.
We can show that $Z[1:l-1]$ is an LCS of $X[1:i-1]$ and $Y[1:j-1]$
including $P[1:k-1]$ as a suffix and excluding $Q[1:r]$ as a subsequence.
Assume by contradiction that there exists a common subsequence $a$ of $X[1:i-1]$ and $Y[1:j-1]$
including $P[1:k-1]$ as a suffix and excluding $Q[1:r]$ as a subsequence, whose length is greater than $l-1$. Then the concatenation of $a$ and $z_l$ will result in a common subsequence of $X[1:i]$ and $Y[1:j]$ including $P[1:k]$ as a suffix and excluding $Q[1:r]$ as a subsequence, whose length is greater than $l$. This is a contradiction.

4. In this case, since $x_i=y_j\neq p_k$, we have $x_i\neq z_l$, otherwise $Z[1:l]$ will not including $P[1:k]$ as a suffix.
Therefore, $Z[1:l]$ must be an LCS of $X[1:i-1]$ and $Y[1:j-1]$ including $P[1:k]$ as a suffix and excluding $Q[1:r]$ as a subsequence, i.e. $Z[1:l]\in W(i-1,j-1,k,r)$.

5. Since $x_i=y_j=q_1$ and $r=1$, we have $x_i\neq z_l$, otherwise $Z[1:l]$ will including $Q[1:r]$ as a subsequence.
Therefore, $Z[1:l]$ must be an LCS of $X[1:i-1]$ and $Y[1:j-1]$ including $P[1:k]$ as a suffix and excluding $Q[1:r]$ as a subsequence, i.e. $Z[1:l]\in W(i-1,j-1,k,r)$.

6. There are two subcases to be distinguished in this case.

6.1. If $z_l=x_i$, then $Z[1:l-1]$ is a common subsequence of $X[1:i-1]$ and $Y[1:j-1]$ excluding $Q[1:r-1]$ as a subsequence. We can show that $Z[1:l-1]$ is an LCS of $X[1:i-1]$ and $Y[1:j-1]$ excluding $Q[1:r-1]$ as a subsequence.
Assume by contradiction that there exists a common subsequence $a$ of $X[1:i-1]$ and $Y[1:j-1]$ excluding $Q[1:r-1]$ as a subsequence, whose length is greater than $l-1$. Then the concatenation of $a$ and $z_l$ will result in a common subsequence of $X[1:i]$ and $Y[1:j]$ excluding $Q[1:r]$ as a subsequence, whose length is greater than $l$. This is a contradiction.
Therefore, in this case we have $Z[1:l-1]\in W(i-1,j-1,k,r-1)$.

6.2. If $z_l\neq x_i$, then $Z[1:l]$ must be an LCS of $X[1:i-1]$ and $Y[1:j-1]$ excluding $Q[1:r]$ as a subsequence, i.e. $Z[1:l]\in W(i-1,j-1,k,r)$.

7. Since $x_i=y_j$ and $r>0, x_i\neq q_r$ or $r=0$, we have $z_l=x_i$, and $Z[1:l-1]$ is a common subsequence of $X[1:i-1]$ and $Y[1:j-1]$ excluding $Q[1:r]$ as a subsequence. We can show that $Z[1:l-1]$ is an LCS of $X[1:i-1]$ and $Y[1:j-1]$ excluding $Q[1:r]$ as a subsequence.
Assume by contradiction that there exists a common subsequence $a$ of $X[1:i-1]$ and $Y[1:j-1]$
excluding $Q[1:r]$ as a subsequence, whose length is greater than $l-1$. Then the concatenation of $a$ and $z_l$ will result in a common subsequence of $X[1:i]$ and $Y[1:j]$ excluding $Q[1:r]$ as a subsequence, whose length is greater than $l$. This is a contradiction.

8. Since $x_i\neq y_j$ and $z_l\neq x_i$, $Z[1:l]$ must be a common subsequence of $X[1:i-1]$ and $Y[1:j]$ including $P[1:k]$ as a suffix and including $Q[1:r]$ as a subsequence. It is obvious that $Z[1:l]$ is also an LCS of $X[1:i-1]$ and $Y[1:j]$ including $P[1:k]$ as a suffix and including $Q[1:r]$ as a subsequence.

9. Since $x_i\neq y_j$ and $z_l\neq y_j$, $Z[1:l]$ must be a common subsequence of $X[1:i]$ and $Y[1:j-1]$ including $P[1:k]$ as a suffix and including $Q[1:r]$ as a subsequence. It is obvious that $Z[1:l]$ is also an LCS of $X[1:i]$ and $Y[1:j-1]$ including $P[1:k]$ as a suffix and including $Q[1:r]$ as a subsequence.

The proof is completed. $\Box$

\section{A simple dynamic programming algorithm}
Our new algorithm for solving the STR-IC-SEQ-EC-LCS problem consists of three main stages. The main idea of the new algorithm can be described by the following Theorem 2.

\begin{theo}\label{th2}
Let $Z[1:l]=z_1,z_2,\cdots,z_l$ be a solution of the STR-IC-SEQ-EC-LCS problem, i.e. $Z[1:l]\in Z(n,m,s,t)$, then its length $l=g(n,m,s,t)$ can be computed by the following formula:

\beql{eq31}
g(n,m,s,t)=\max_{1\leq i\leq n,1\leq j\leq m,1\leq r\leq t}\left\{f(i,j,s,r)+h(i+1,j+1,r)\right\}
\eeq

where $f(i,j,s,r)$ is the length of an LCS in $W(i,j,s,r)$ defined by Definiton 2, and $h(i,j,r)$ is the length of an LCS in $U(i,j,r)$ defined by Definiton 3.

\end{theo}
\noindent{\bf Proof.}

Since $Z[1:l]\in Z(n,m,s,t)$, $Z[1:l]$ must be an LCS of $X$ and $Y$ including $P$ as a substring, and excludes $Q$ as a subsequence.
Let the first appearance of the string $P$ in $Z[1:l]$ starts from position $l'-s+1$ to $l'$ for some positive integer $s\leq l'\leq l$, i.e. $Z[l'-s+1:l']=P$.

Let
\[
r^*=\max_{1\leq r\leq t}\left\{r| Q[1:r]\texttt{ is a subsequence of }Z[1:l']\right\}
\]

Since $Z[1:l']$ excludes $Q$ as a subsequence, we have $r^*<t$, and thus $Z[1:l']$ excludes $Q[1:r^*+1]$ as a subsequence.
For the same reason, $Z[l'+1:l]$ excludes $Q[r^*+1:t]$ as a subsequence.

Let
\[
(i^*,j^*)=\min_{1\leq i\leq n,1\leq j\leq m}\left\{(i,j)| Z[1:l']\texttt{ is a common subsequence of }X[1:i]\texttt{ and }Y[1:j]\right\}
\]

Then, $Z[1:l']$ is a common subsequence of $X[1:i^*]$ and $Y[1:j^*]$ including $P$ as a suffix and excluding $Q[1:r^*+1]$ as a subsequence. It follows from Definition 2 that

\beql{eq32}
l'\leq f(i^*,j^*,s,r^*+1)
\eeq

Since $Z[1:l]$ is a common subsequence of $X$ and $Y$, $Z[l'+1:l]$ must be a common subsequence of $X[i^*+1:n]$ and $Y[j^*+1:m]$. We have known $Z[l'+1:l]$ excludes $Q[r^*+1:t]$ as a subsequence. Therefore, $Z[l'+1:l]$ is a common subsequence of $X[i^*+1:n]$ and $Y[j^*+1:m]$ excluding $Q[r^*+1:t]$ as a subsequence.
It follows from Definition 3 that

\beql{eq33}
l-l'\leq h(i^*+1,j^*+1,r^*+1)
\eeq

Combining formulas \eqn{eq32} and \eqn{eq33} we have,

\[
l\leq f(i^*,j^*,s,r^*+1)+h(i^*+1,j^*+1,r^*+1)
\]

Therefore,

\beql{eq34}
l\leq \max_{1\leq i\leq n,1\leq j\leq m,1\leq r\leq t}\left\{f(i,j,s,r)+h(i+1,j+1,r)\right\}
\eeq

On the other hand, for any $a\in W(i,j,s,r)$ and $b\in U(i+1,j+1,r)$, $1\leq i\leq n,1\leq j\leq m,1\leq r\leq t$, then $c=a\bigoplus b$, the concatenation of $a$ and $b$, must be a common subsequence of $X[1:n]$ and $Y[1:m]$ including $P$ as a substring. Furthermore, we can prove $c$ excludes $Q$ as a subsequence.

In fact, let
\[
r^*=\max_{0\leq r'\leq t}\left\{r'| Q[1:r']\texttt{ is a subsequence of }a\right\}
\]

We then have $r*<r$, since $a$ excludes $Q[1:r]$ as a subsequence.

In this case, if $c$ includes $Q$ as a subsequence, then $b$ must include $Q[r^*+1:t]$ as a subsequence. It follows from $r^*+1\leq r$ that $b$ includes $Q[r:t]$ as a subsequence. This is a contradiction.

Therefore, we have $c=a\bigoplus b$ is a common subsequence of $X[1:n]$ and $Y[1:m]$ including $P$ as a substring and excluding $Q$ as a subsequence, and thus $|a\bigoplus b|\leq l$. That is:
\beql{eq35}
\max_{1\leq i\leq n,1\leq j\leq m,1\leq r\leq t}\left\{f(i,j,s,r)+h(i+1,j+1,r)\right\}\leq l
\eeq

Combining formulas \eqn{eq34} and \eqn{eq35} we have,
\[
l=\max_{1\leq i\leq n,1\leq j\leq m,1\leq r\leq t}\left\{f(i,j,s,r)+h(i+1,j+1,r)\right\}
\]

The proof is completed. $\Box$

The first stage is to find LCSs in $W(i,j,k,r)$.
Let $f(i,j,k,r)$ denote the length of an LCS in $W(i,j,k,r)$. By the optimal substructure properties of the STR-IC-SEQ-EC-LCS problem shown in Theorem 1, we can build the following recursive formula for computing $f(i,j,k,r)$.
For any $1\leq i\leq n, 1\leq j\leq m, 0\leq k\leq s$, and $0\leq r\leq t$, the values of $f(i,j,k,r)$ can be computed by the following recursive formula \eqn{eq36}.

\beql{eq36}
f(i,j,k,r)=\left\{\begin{array}{ll}
\max\left\{
f(i-1,j,k,r),f(i,j-1,k,r)
\right\} & \textrm{if } x_i\neq y_j\\
1+f(i-1,j-1,k-1,r) & \textrm{if } x_i=y_j=p_k\wedge (r=0 \vee x_i\neq q_r)\\
f(i-1,j-1,k,r) & \textrm{if } x_i=y_j=p_k=q_r\wedge r=1\\
\max\left\{
1+f(i-1,j-1,k-1,r-1),f(i-1,j-1,k,r)
\right\} & \textrm{if } x_i=y_j=p_k=q_r\wedge r>1\\

f(i-1,j-1,k,r) & \textrm{if } i,j,k>0\wedge x_i=y_j\neq p_k\\

1+f(i-1,j-1,k,r) & \textrm{if } k=0\wedge x_i=y_j\wedge (r=0 \vee x_i\neq q_r)\\
f(i-1,j-1,k,r) & \textrm{if } k=0\wedge x_i=y_j\wedge (r=1 \wedge x_i=q_r)\\
\max\left\{
1+f(i-1,j-1,k,r-1),f(i-1,j-1,k,r)
\right\} & \textrm{if } k=0\wedge x_i=y_j=q_r\wedge r>1\\
\end{array} \right.
\eeq

The boundary conditions of this recursive formula are $f(i,0,0,0)=f(0,j,0,0)=0$ and $f(i,0,k,r)=f(0,j,k,r)=-\infty$ for any $0\leq i\leq n, 0\leq j\leq m, 1\leq k\leq s$, and $1\leq r\leq t$.

Based on this formula, our algorithm for computing $f(i,j,k,r)$ is a standard dynamic programming algorithm. By the recursive formula \eqn{eq31}, the dynamic programming algorithm for computing $f(i,j,k,r)$ can be implemented as the following Algorithm 1.

\begin{algorithm}
\caption{Suffix}
{\bf Input:} Strings $X=x_1\cdots x_n$, $Y=y_1\cdots y_m$ of lengths $n$ and $m$, respectively, and two constrained sequences $P=p_1p_2\cdots p_s$ and $Q=q_1q_2\cdots q_t$ of lengths $s$ and $t$\\
{\bf Output:} $f(i,j,k,r)$, the length of an LCS of $X[1:i]$ and $Y[1:j]$ including $P[1:k]$ as a suffix, and excluding $Q[1:r]$ as a subsequence, for all $1\leq i\leq n, 1\leq j\leq m, 0\leq k\leq s$, and $0\leq r\leq t$.
\begin{algorithmic}[1]
\FORALL{$i,j,k,r$ , $0\leq i\leq n, 0\leq j\leq m, 0\leq k \leq s$ and $0\leq r\leq t$}
\STATE $f(i,0,k,r),f(0,j,k,r)\leftarrow -\infty, f(i,0,0,0),f(0,j,0,0)\leftarrow 0$ \{boundary condition\}
\ENDFOR
\FORALL{$i,j,k,r$ , $1\leq i\leq n, 1\leq j\leq m, 0\leq k \leq s$ and $0\leq r\leq t$}
\IF {$x_i\neq y_j$}
\STATE $f(i,j,k,r) \leftarrow \max\{f(i-1,j,k,r),f(i,j-1,k,r)\}$\\
\ELSIF {$k>0$ \AND $x_i=p_k$}

\IF {$r=0$ \AND $x_i\neq q_r$}
\STATE $f(i,j,k,r) \leftarrow 1+f(i-1,j-1,k-1,r)$\\

\ELSIF {$r=1$ \AND $x_i=q_r$}
\STATE $f(i,j,k,r) \leftarrow f(i-1,j-1,k,r)$\\
\ELSE
\STATE $f(i,j,k,r) \leftarrow \max\{1+f(i-1,j-1,k-1,r-1),f(i-1,j-1,k,r)\}$\\
\ENDIF
\ELSIF {$k=0$}

\IF {$r=0$ \OR $x_i\neq q_r$}
\STATE $f(i,j,k,r) \leftarrow 1+f(i-1,j-1,k,r)$\\

\ELSIF {$r=1$ \AND $x_i=q_r$}
\STATE $f(i,j,k,r) \leftarrow f(i-1,j-1,k,r)$\\

\ELSE
\STATE $f(i,j,k,r) \leftarrow \max\{1+f(i-1,j-1,k,r-1),f(i-1,j-1,k,r)\}$\\

\ENDIF
\ELSE

\STATE $f(i,j,k,r) \leftarrow f(i-1,j-1,k,r)$\\
\ENDIF
\ENDFOR
\end{algorithmic}
\end{algorithm}

It is obvious that the algorithm requires $O(nmst)$ time and space.
For each value of $f(i,j,k,r)$ computed by algorithm $\emph{Suffix}$, the corresponding LCS of $X[1:i]$ and $Y[1:j]$ including $P[1:k]$ as a subsequence, and including $Q[1:r]$ as a suffix, can be constructed by backtracking through the computation paths from $(i,j,k,r)$ to $(0,0,0,0)$.
The following algorithm $back(i,j,k,r)$ is the backtracking algorithm to obtain the LCS, not only its length. The time complexity of the algorithm $back(i,j,k,r)$ is obviously $O(n+m)$.

\begin{algorithm}
\caption{$back(i,j,k,r)$}
{\bf Input:} Integers $i,j,k,r$\\
{\bf Output:} The LCS of $X[1:i]$ and $Y[1:j]$ including $P[1:k]$ as a suffix and excluding $Q[1:r]$ as a subsequence
\begin{algorithmic}[1]
\IF {$i<1$ \OR $j<1$}
\RETURN
\ENDIF
\IF {$x_i\neq y_j$}
\IF {$f(i-1,j,k,r)>f(i,j-1,k,r)$}
\STATE $back(i-1,j,k,r)$\\
\ELSE
\STATE $back(i,j-1,k,r)$\\
\ENDIF

\ELSIF {$k>0$ \AND $x_i=p_k$}

\IF {$r=0$ \AND $x_i\neq q_r$}
\STATE $back(i-1,j-1,k-1,r)$\\
\PRINT $x_i$\\

\ELSIF {$r=1$ \AND $x_i=q_r$}
\STATE $back(i-1,j-1,k,r)$\\
\ELSE
\IF {$1+f(i-1,j-1,k-1,r-1)>f(i-1,j-1,k,r)$}
\STATE $back(i-1,j-1,k-1,r-1)$\\
\PRINT $x_i$\\
\ELSE
\STATE $back(i-1,j-1,k,r)$\\
\ENDIF
\ENDIF

\ELSIF {$k=0$}

\IF {$r=0$ \OR $x_i\neq q_r$}
\STATE $back(i-1,j-1,k,r)$\\
\PRINT $x_i$\\

\ELSIF {$r=1$ \AND $x_i=q_r$}
\STATE $back(i-1,j-1,k,r)$\\
\ELSE
\IF {$1+f(i-1,j-1,k,r-1)>f(i-1,j-1,k,r)$}
\STATE $back(i-1,j-1,k,r-1)$\\
\PRINT $x_i$\\
\ELSE
\STATE $back(i-1,j-1,k,r)$\\
\ENDIF
\ENDIF
\ELSE
\STATE $back(i-1,j-1,k,r)$\\
\ENDIF

\end{algorithmic}
\end{algorithm}

The second stage of our algorithm is to find LCSs in $U(i,j,k)$. The length of an LCS in $U(i,j,k)$ is denoted as $h(i,j,k)$.
Chen et al.\cite{1} presented a dynamic programming algorithm with $O(nmt)$ time and space. A reverse version of the dynamic programming algorithm for computing $h(i,j,k)$ can be described as follows.

\begin{algorithm}
\caption{SEQ-EC-R}
{\bf Input:} Strings $X=x_1\cdots x_n$, $Y=y_1\cdots y_m$ of lengths $n$ and $m$, respectively, and a constrained sequence $Q=q_1q_2\cdots q_t$ of lengths $t$\\
{\bf Output:} $h(i,j,k)$, the length of an LCS of $X[i:n]$ and $Y[j:m]$ excluding $Q[k:t]$ as a subsequence,
 for all $1\leq i\leq n, 1\leq j\leq m, 0\leq k\leq t$.
\begin{algorithmic}[1]
\FORALL{$i,j,k$ , $0\leq i\leq n, 0\leq j\leq m, 1\leq k \leq t$}
\STATE $h(i,m+1,k),h(n+1,j,k)\leftarrow -\infty$ \{boundary condition\}
\ENDFOR
\FOR{$i=n$ down to $1$}
\FOR{$j=m$ down to $1$}
\FOR{$k=t+1$ down to $1$}
\IF {$x_i\neq y_j$}
\STATE $h(i,j,k) \leftarrow \max\{h(i+1,j,k),h(i,j+1,k)\}$\\
\ELSE
\IF {$k>t$ \OR $k\leq t$ \AND $x_i\neq q_k$}
\STATE $h(i,j,k) \leftarrow 1+h(i+1,j+1,k)$\\
\ELSIF {$x_i=q_k$}

\IF {$k=t$}
\STATE $h(i,j,k) \leftarrow h(i+1,j+1,k)$\\
\ELSE
\STATE $h(i,j,k) \leftarrow \max\{1+h(i+1,j+1,k+1),h(i+1,j+1,k)\}$\\
\ENDIF
\ENDIF
\ENDIF
\ENDFOR
\ENDFOR
\ENDFOR
\end{algorithmic}
\end{algorithm}

For each value of $h(i,j,k)$ computed by algorithm $\emph{SEQ-EC-R}$, the corresponding LCS of $X[i:n]$ and $Y[j:m]$ excluding $Q[k:t]$ as a subsequence, can be constructed by backtracking through the computation paths from $(i,j,k)$ to $(0,0,0)$.
The following algorithm $backr(i,j,k)$ is the backtracking algorithm to obtain the corresponding LCS, not only its length. The time complexity of the algorithm $backr(i,j,k)$ is obviously $O(n+m)$.

\begin{algorithm}
\caption{$backr(i,j,k)$}
{\bf Input:} Integers $i,j,k$\\
{\bf Output:} The LCS of $X[i:n]$ and $Y[j:m]$ including $P[k:s]$ as a subsequence
\begin{algorithmic}[1]
\IF {$i>n$ \OR $j>m$}
\RETURN
\ENDIF
\IF {$x_i\neq y_j$}
\IF {$h(i+1,j,k)>h(i,j+1,k)$}
\STATE $backr(i+1,j,k)$\\
\ELSE
\STATE $backr(i,j+1,k)$\\
\ENDIF
\ELSE
\IF {$k>t$ \OR $k\leq t$ \AND $x_i\neq q_k$}
\PRINT $x_i$\\
\STATE $backr(i+1,j+1,k)$\\
\ELSIF {$x_i=q_k$}

\IF {$k=t$}
\STATE $backr(i+1,j+1,k)$\\
\ELSE

\IF {$h(i+1,j+1,k)>1+h(i+1,j+1,k+1)$}
\STATE $backr(i+1,j+1,k)$\\
\ELSE
\PRINT $x_i$\\
\STATE $backr(i+1,j+1,k+1)$\\
\ENDIF
\ENDIF
\ENDIF
\ENDIF
\end{algorithmic}
\end{algorithm}

By Theorem 2, the dynamic programming matrices $f(i,j,k,r)$ and $h(i,j,k)$ computed by the algorithms $\emph{Suffix}$ and $\emph{SEQ-EC-R}$ can now be combined to obtain the solutions of the STR-IC-SEQ-EC-LCS problem as follows. This is the final stage of our algorithm.

\begin{algorithm}
\caption{STR-IC-SEQ-EC-LCS}
{\bf Input:} Strings $X=x_1\cdots x_n$, $Y=y_1\cdots y_m$ of lengths $n$ and $m$, respectively, and two constrained sequences $P=p_1p_2\cdots p_s$ and $Q=q_1q_2\cdots q_t$ of lengths $s$ and $t$\\
{\bf Output:} The constrained LCS of $X$ and $Y$ including $P$ as a substring, and including $Q$ as a subsequence.
\begin{algorithmic}[1]
\STATE Suffix \COMMENT{compute $f(i,j,k,r)$}\\
\STATE SEQ-EC-R \COMMENT{compute $h(i,j,k)$}\\
\STATE $i^*,j^*,k^* \leftarrow 0, tmp \leftarrow -\infty$\\
\FOR{$i=1$ to $n$}
\FOR{$j=1$ to $m$}
\FOR{$k=1$ to $t$}
\STATE $x \leftarrow f(i,j,s,k)+h(i+1,j+1,k)$\\
\IF{$tmp<x$}
\STATE $tmp \leftarrow x, i^* \leftarrow i, j^* \leftarrow j, k^* \leftarrow k$\\
\ENDIF
\ENDFOR
\ENDFOR
\ENDFOR
\IF{$tmp>0$}
\STATE $back(i^*,j^*,s,k^*)$\\
\STATE $backr(i^*+1,j^*+1,k^*)$\\
\ENDIF
\RETURN $\max\{0,tmp\},i^*,j^*,k^*$
\end{algorithmic}
\end{algorithm}

From the 'for' loops of the algorithm, it is readily seen that the algorithm requires $O(nmt)$ time. Therefore, the overall time of our algorithm for solving the STR-IC-SEQ-EC-LCS problem is $O(nmst)$.

\section{Improvements of the algorithm}

S. Deorowicz\cite{3} proposed the first quadratic-time algorithm for the STR-IC-LCS problem. A similar idea can be exploited to improve the time complexity of our dynamic programming algorithm for solving the STR-IC-SEQ-EC-LCS problem.
The improved algorithm is also based on dynamic programming with some preprocessing. To show its correctness it is necessary to prove some more structural properties of the problem.

Let $Z[1:l]=z_1,z_2,\cdots,z_l\in Z(n,m,s,t)$, be a constrained LCS of $X$ and $Y$ including $P$ as a substring and excluding $Q$ as a subsequence.
Let also $I=(i_1,j_1),(i_2,j_2),\cdots,(i_l,j_l)$ be a sequence of indices of $X$ and $Y$ such that
$Z[1:l]=x_{i_1},x_{i_2},\cdots,x_{i_l}$ and $Z[1:l]=y_{j_1},y_{j_2},\cdots,y_{j_l}$. From the problem statement, there must exist an index $d\in [1,l-t+1]$ such that $P=x_{i_d},x_{i_{d+1}},\cdots,x_{i_{d+s-1}}$ and $P=y_{j_d},y_{j_{d+1}},\cdots,y_{j_{d+s-1}}$.

\begin{theo}\label{th3}
Let $i'_d=i_d$ and for all $e\in[1,s-1]$, $i'_{d+e}$ be the smallest possible, but larger than $i'_{d+e-1}$, index of $X$ such that $x_{i_{d+e}}=x_{i'_{d+e}}$.
The sequence of indices

$I'=(i_1,j_1),(i_2,j_2),\cdots,(i_{d-1},j_{d-1}),(i'_{d},j_{d}),(i'_{d+1},j_{d+1}),\cdots,(i'_{d+s-1},j_{d+s-1}),(i_{d+s},j_{d+s}),\cdots,(i_l,j_l)$

defines the same constrained LCS as $Z[1:l]$.
\end{theo}

\noindent{\bf Proof.}

From the definition of indices $i'_{d+e}$, it is obvious that they form an increasing sequence, since $i'_d=i_d$, and $i'_{d+s-1}\leq i_{d+s-1}$.
The sequence $i'_d,\cdots,i'_{d+s-1}$ is of course a compact appearance of $P$ in $X$ starting at $i_d$. Therefore,
both components of $I'$ pairs form increasing sequences and for any $(i'_u,j_u)$, $x_{i'_u}=y_{j_u}$. Therefore,
$I'$ defines the same constrained LCS as $Z[1:l]$.

The proof is completed. $\Box$

The same property is also true for the $j$th components of the sequence $I$. Therefore, we can conclude that when finding a constrained LCS in $Z(i,j,k,r)$, instead of checking any common subsequences of $X$ and $Y$ it suffices to check only such common subsequences that contain compact appearances of $P$ both in $X$ and $Y$.
The number of different compact appearances of $Q$ in $X$ and $Y$ will be denoted by $\delta_x$ and $\delta_y$, respectively. It is obvious that $\delta_x\delta_y\leq\delta$, since a pair $(i,j)$ defines a compact appearance of $Q$ in $X$ starting at $i$th position and compact appearance of $Q$ in $Y$ starting at $j$th position only for some matches.

Base on Theorem 2, we can reduce the time complexity of our dynamic programming from $O(nmst)$ to $O(nmt)$.
The improved algorithm consists of also three main stages.

\begin{defi}\label{df5}
For each occurrence $i$ of the first character $p_1$ of $P[1:s]$ in $X[1:n]$, $lx_i$ is defined as the index of the last character $p_s$ of a compact appearance of $P$ in $X$. If $x_i\neq p_1$ or there is no compact appearance of $P$ after $i$, then $lx_i=0$. Similarly, for each occurrence $j$ of the first character $p_1$ of $P[1:s]$ in $Y[1:m]$, $ly_j$ is defined as the index of the last character $p_s$ of a compact appearance of $P$ in $Y$.
\end{defi}

In the first stage both sequences $X$ and $Y$ are preprocessed to determine two corresponding arrays $lx$ and $ly$.

\begin{algorithm}
\caption{Prep}
{\bf Input:} $X,Y$\\
{\bf Output:} For each $1\leq i\leq n$, the minimal index $r=lx_i$ such that $X[i:r]$ includes $P$ as a subsequence\\
For each $1\leq j\leq m$, the minimal index $r=ly_j$ such that $Y[j:r]$ includes $P$ as a subsequence
\begin{algorithmic}[1]
\FOR{$i=1$ to $n$}
\IF{$x_i=p_1$}
\STATE $lx_i \leftarrow left(X,n,i)$\\
\ELSE
\STATE $lx_i \leftarrow 0$\\
\ENDIF
\ENDFOR
\FOR{$j=1$ to $m$}
\IF{$y_j=p_1$}
\STATE $ly_j \leftarrow left(Y,m,j)$\\
\ELSE
\STATE $ly_j \leftarrow 0$\\
\ENDIF
\ENDFOR
\end{algorithmic}
\end{algorithm}

In the algorithm Prep, function $left$ is used to find the index $lx_i$ of the last character $p_s$ of a compact appearance of $P$.

\begin{algorithm}
\caption{$left(X,n,i)$}
{\bf Input:} Integers $n,i$ and $X[1:n]$\\
{\bf Output:} The minimal index $r$ such that $X[i:r]$ includes $P$ as a subsequence
\begin{algorithmic}[1]
\STATE $a \leftarrow i+1, b \leftarrow 2$\\
\WHILE{$a \leq n$ \AND $b\leq s$}
\IF{$x_a=p_b$}
\STATE $b \leftarrow b+1$\\
\ELSE
\STATE $a \leftarrow a+1$\\
\ENDIF
\ENDWHILE
\IF{$b>s$}
\RETURN{$a-1$}
\ELSE
\RETURN{0}
\ENDIF
\end{algorithmic}
\end{algorithm}

In the second stage two DP matrices of SEQ-EC-LCS problem are computed: $h(i,j,k)$, the reverse one defined by Definition 3, and $v(i,j,k)$, the forward one defined by Definition 4. Both of the DP matrices can be computed by the SEQ-EC-LCS algorithm of Chen et al.\cite{1}.

\begin{algorithm}
\caption{SEQ-EC}
{\bf Input:} Strings $X=x_1\cdots x_n$, $Y=y_1\cdots y_m$ of lengths $n$ and $m$, respectively, and a constrained sequence $Q=q_1q_2\cdots q_t$ of length $t$\\
{\bf Output:} $v(i,j,k)$, the length of an LCS of $X[1:i]$ and $Y[1:j]$ excluding $Q[1:k]$ as a subsequence,
 for all $1\leq i\leq n, 1\leq j\leq m, 0\leq k\leq t$.
\begin{algorithmic}[1]
\FORALL{$i,j,k$ , $0\leq i\leq n, 0\leq j\leq m, 1\leq k \leq t$}
\STATE $h(i,0,k),h(0,j,k)\leftarrow -\infty$ \{boundary condition\}
\ENDFOR
\FOR{$i=1$ to $n$}
\FOR{$j=1$ to $m$}
\FOR{$k=0$ to $t$}
\IF {$x_i\neq y_j$}
\STATE $v(i,j,k) \leftarrow \max\{v(i-1,j,k),v(i,j-1,k)\}$\\
\ELSE
\IF {$k=0$ \OR $k>0$ \AND $x_i\neq q_k$}
\STATE $v(i,j,k) \leftarrow 1+v(i-1,j-1,k)$\\
\ELSIF {$x_i=q_k$}

\IF {$k=1$}
\STATE $v(i,j,k) \leftarrow v(i-1,j-1,k)$\\
\ELSE
\STATE $v(i,j,k) \leftarrow \max\{1+v(i-1,j-1,k-1),v(i-1,j-1,k)\}$\\
\ENDIF
\ENDIF
\ENDIF
\ENDFOR
\ENDFOR
\ENDFOR
\end{algorithmic}
\end{algorithm}

\begin{algorithm}
\caption{$backf(i,j,k)$}
{\bf Input:} Integers $i,j,k$\\
{\bf Output:} The LCS of $X[1:i]$ and $Y[1:j]$ excluding $Q[1:k]$ as a subsequence
\begin{algorithmic}[1]
\IF {$i<1$ \OR $j<1$}
\RETURN
\ENDIF
\IF {$x_i\neq y_j$}
\IF {$v(i-1,j,k)>v(i,j-1,k)$}
\STATE $backr(i-1,j,k)$\\
\ELSE
\STATE $backr(i,j-1,k)$\\
\ENDIF
\ELSE
\IF {$k=0$ \OR $k>0$ \AND $x_i\neq q_k$}
\STATE $backr(i-1,j-1,k)$\\
\PRINT $x_i$\\
\ELSIF {$x_i=q_k$}

\IF {$k=1$ \OR $v(i-1,j-1,k)>1+v(i-1,j-1,k-1)$}
\STATE $backr(i-1,j-1,k)$\\
\ELSE
\STATE $backr(i-1,j-1,k-1)$\\
\PRINT $x_i$\\
\ENDIF
\ENDIF
\ENDIF
\end{algorithmic}
\end{algorithm}

In the last stage, two preprocessed arrays $lx$ and $ly$ are used to determine the final results.
To this end for each match $(i,j)$ for $X$ and $Y$ the ends $(lx_i,ly_i)$ of compact appearances of $P$ in $X$ starting at position $i$ and in $Y$ starting at position $j$ are read.
The length of an STR-IC-SEQ-EC-LCS, $g(n,m,s,t)$ defined by Definition 1, containing these appearances of $P$ is determined as a sum of three parts.
For some indices $i,j,k,r$, $v(i-1,j-1,k)$, the constrained LCS length of prefixes of $X$ and $Y$ ending at positions $i-1$ and $j-1$, excluding $Q[1:k]$ as a subsequence, $h(lx_i+1,ly_j+1,r)$ the constrained LCS length of suffixes of $X$ and $Y$ starting at positions $lx_i+1$ and $ly_j+1$, excluding $Q[r:t]$ as a subsequence, and the constraint length $s$.
The integers $k$ and $r$ have some relations.

\begin{defi}\label{df6}
For each integer $k,1\leq k\leq t$, the index $\alpha(k)$ is defined as:
\beql{eq37}
\alpha(k)=\max_{0\leq r\leq s-k+1}\{r|P\texttt{ includes }Q[k:k+r-1]\texttt{ as a subsequence}\}
\eeq
\end{defi}

Since the constrained LCS $A$ of prefixes of $X$ and $Y$ ending at positions $i-1$ and $j-1$, excludes $Q[1:k]$ as a subsequence, the concatenation of $A$ and $P$ will exclude $Q[1:r]$ as a subsequence, where $r=k+\alpha(k)$.
The constrained LCS $B$ of suffixes of $X$ and $Y$ starting at positions $lx_i+1$ and $ly_j+1$, excludes $Q[r:t]$ as a subsequence. Therefore, the concatenation of $A$,$P$ and $B$ excludes $Q$ as a subsequence.

According to the matrices $v(i,j,k)$ and $h(i,j,k)$, backtracking can be used to obtain the optimal subsequence, not only its length.

\begin{algorithm}
\caption{$\alpha(k)$}
{\bf Input:} Integers $k$\\
{\bf Output:} The maximum length $r$ $(0\leq r\leq s-k+1)$ such that $P$ includes $P[k:k+r-1]$ as a subsequence
\begin{algorithmic}[1]
\STATE $a \leftarrow k, b \leftarrow 1, r \leftarrow 0$\\
\WHILE{$a \leq s$ \AND $b\leq t$}
\IF{$p_a=q_b$}
\STATE $a \leftarrow a+1, r \leftarrow r+1$\\
\ELSE
\STATE $b \leftarrow b+1$\\
\ENDIF
\ENDWHILE
\RETURN{$r$}
\end{algorithmic}
\end{algorithm}

\begin{algorithm}
\caption{STR-IC-SEQ-EC-LCS}
{\bf Input:} Strings $X=x_1\cdots x_n$, $Y=y_1\cdots y_m$ of lengths $n$ and $m$, respectively, and two constrained sequences $P=p_1p_2\cdots p_s$ and $Q=q_1q_2\cdots q_t$ of lengths $s$ and $t$\\
{\bf Output:} The length of an LCS of $X$ and $Y$ including $P$ as a substring, and excluding $Q$ as a subsequence.
\begin{algorithmic}[1]
\STATE SEQ-EC \COMMENT{compute $v(i,j,k)$}\\
\STATE SEQ-EC-R \COMMENT{compute $h(i,j,k)$}\\
\STATE Prep \COMMENT{compute $lx,ly$}\\
\STATE $i^*,j^*,k^*,r^*\leftarrow 0, tmp \leftarrow 0$\\
\FOR{$i=1$ \TO $n$}
\FOR{$j=1$ \TO $m$}
\IF{$lx_i>0$ \AND $ly_j>0$}
\FOR{$k=1$ \TO $t$}
\STATE $r \leftarrow k+\alpha(k)$\\
\STATE $c \leftarrow v(i-1,j-1,k)+h(lx_i+1,ly_j+1,r)+s$\\

\IF{$r>t$}
\STATE $tmp \leftarrow \infty$\\
\ENDIF
\IF{$tmp<c$}
\STATE $tmp \leftarrow c, i^* \leftarrow i, j^* \leftarrow j, k^* \leftarrow k, r^* \leftarrow r$\\
\ENDIF
\ENDFOR
\ENDIF
\ENDFOR
\ENDFOR
\IF{$tmp>0$}
\STATE $backf(i^*-1,j^*-1,k^*)$\\
\PRINT $P$
\STATE $backr(lx_{i^*}+1,ly_{j^*}+1,r^*)$\\
\ENDIF
\RETURN $\max\{0,tmp\},i^*,j^*,k^*,r^*$
\end{algorithmic}
\end{algorithm}

\begin{theo}\label{th4}
The algorithm STR-IC-SEQ-EC-LCS correctly computes a constrained LCS in $Z(n,m,s,t)$. The algorithm requires $O(nmt)$ time and to $O(nmt)$ space in the worst case.
\end{theo}

\noindent{\bf Proof.}

Let $Z[1:l]=z_1,z_2,\cdots,z_l$ be a solution of the STR-IC-SEQ-EC-LCS problem, i.e. $Z[1:l]\in Z(n,m,s,t)$, and its length be denoted as $l=g(n,m,s,t)$.
To prove the theorem, we have to prove in fact that

\beql{eq38}
g(n,m,s,t)=s+\max_{1\leq i\leq n,1\leq j\leq m,0\leq k\leq t}\left\{v(i-1,j-1,k)+h(lx_i+1,ly_j+1,k+\alpha(k))\right\}
\eeq

where $h(i,j,k)$ is the length of an LCS in $U(i,j,k)$ defined by Definiton 3, and $v(i,j,k)$ is the length of an LCS in $V(i,j,k)$ defined by Definiton 4.

Since $Z[1:l]\in Z(n,m,s,t)$, $Z[1:l]$ must be an LCS of $X$ and $Y$ including $P$ as a substring, and excludes $Q$ as a subsequence.
Let the first appearance of the string $P$ in $Z[1:l]$ starts from position $l'-s+1$ to $l'$ for some positive integer $s\leq l'\leq l$, i.e. $Z[l'-s+1:l']=P$.

Let
\[
r^*=\max_{1\leq r\leq t}\left\{r| Q[1:r]\texttt{ is a subsequence of }Z[1:l'-s]\right\}
\]

Since $Z[1:l'-s]$ excludes $Q$ as a subsequence, we have $r^*<t$, and thus $Z[1:l'-s]$ excludes $Q[1:r^*+1]$ as a subsequence.

Let
\[
(i^*,j^*)=\min_{1\leq i\leq n,1\leq j\leq m}\left\{(i,j)| Z[1:l'-s+1]\texttt{ is a common subsequence of }X[1:i]\texttt{ and }Y[1:j]\right\}
\]

Then, $x_{i^*}=y_{j^*}=p_1=z_{l'-s+1}$, and $x_{lx_{i^*}}=y_{ly_{j^*}}=p_s=z_{l'}$.

Therefore, $Z[1:l'-s]$ is a common subsequence of $X[1:i^*-1]$ and $Y[1:j^*-1]$ excluding $Q[1:r^*+1]$ as a subsequence; $Z[l'+1:l]$ is a common subsequence of $X[lx_{i^*}+1:n]$ and $Y[ly_{j^*}+1:m]$.

It follows from Definition 4 that

\beql{eq39}
l'-s\leq v(i^*-1,j^*-1,r^*+1)
\eeq

Since $Q[1:r^*]$ is the longest prefix of $Q$ in $Z[1:l'-s]$, and $$\alpha(r^*+1)=\max_{0\leq r\leq s-r^*+2}\{r|P\texttt{ includes }Q[r^*+1:r^*+r]\texttt{ as a subsequence}\}$$
we have, $Z[1:l']$ includes $Q[1:r^*+\alpha(r^*+1)]$ as a subsequence.
It follows from $Z[1:l]$ excludes $Q$ as a subsequence that
$Z[l'+1:l]$ excludes $Q[r^*+1+\alpha(r^*+1):t]$ as a subsequence.
Therefore, we have $Z[l'+1:l]$ is a common subsequence of $X[lx_{i^*}+1:n]$ and $Y[ly_{j^*}+1:m]$ excluding $Q[r^*+1+\alpha(r^*+1):t]$ as a subsequence. It follows from Definition 3 that
\beql{eq310}
l-l'\leq h(lx_{i^*}+1,ly_{j^*}+1,r^*+1+\alpha(r^*+1))
\eeq

Combining formulas \eqn{eq39} and \eqn{eq310} we have,

\[
l-s\leq v(i^*-1,j^*-1,r^*+1)+h(lx_{i^*}+1,ly_{j^*}+1,r^*+1+\alpha(r^*+1))
\]

Therefore,

\beql{eq311}
l\leq s+\max_{1\leq i\leq n,1\leq j\leq m,0\leq k\leq t}\left\{v(i-1,j-1,k)+h(lx_i+1,ly_j+1,k+\alpha(k))\right\}
\eeq

On the other hand, for any $a\in V(i,j,k)$ and $b\in U(lx_i+1,ly_j+1,k+\alpha(k))$, $1\leq i\leq n,1\leq j\leq m,1\leq k\leq t$, let $c=a\bigoplus P \bigoplus b$. If $lx_i>0$ and $ly_j>0$, then $c$ must be a common subsequence of $X[1:n]$ and $Y[1:m]$ including $P$ as a substring. Furthermore, we can prove $c$ excludes $Q$ as a subsequence.

In fact, since $a$ excludes $Q[1:k]$ as a subsequence, the length of the longest prefix of $Q$ in $a$ is at most $k-1$, and thus the length of the longest prefix of $Q$ in $a\bigoplus P$ is at most $k-1+\alpha(k)$. Since $b$ is a common subsequence of $X[lx_i+1:n]$ and $Y[ly_j+1:m]$ excluding $Q[k+\alpha(k):t]$ as a subsequence, we have, $c=a\bigoplus P \bigoplus b$ is a common subsequence of $X[1:n]$ and $Y[1:m]$ including $P$ as a substring and excluding $Q$ as a subsequence, and thus $|a\bigoplus P\bigoplus b|\leq l$.
Therefore,
\beql{eq312}
s+\max_{1\leq i\leq n,1\leq j\leq m,0\leq k\leq t}\left\{v(i-1,j-1,k)+h(lx_i+1,ly_j+1,k+\alpha(k))\right\}\leq l
\eeq

Combining formulas \eqn{eq311} and \eqn{eq312} we have,
\[
l=\max_{1\leq i\leq n,1\leq j\leq m,0\leq k\leq t}\left\{v(i-1,j-1,k)+h(lx_i+1,ly_j+1,k+\alpha(k))\right\}
\]

The time and space complexities of the algorithm are dominated by the computation of the two dynamic programming matrices $v(i,j,k)$ and $h(i,j,k)$.
It is obvious that they are all $O(nmt)$ in the worst case.

The proof is completed. $\Box$

\section{Concluding remarks}
We have suggested a new dynamic programming solution for the new generalized constrained longest common subsequence problem STR-IC-SEQ-EC-LCS. The first dynamic programming algorithm requires $O(nmst)$ in the worst case, where $n,m,s,t$ are the lengths of the four input sequences respectively. The time complexity can be reduced further to cubic time in a more detailed analysis.
Many other generalized constrained longest common subsequence (GC-LCS) problems have similar structures.
It is not clear that whether the same technique of this paper can be applied to these problems to achieve efficient algorithms. We will investigate these problems further.

\end{document}